\documentclass{article}

\usepackage{amsmath,amssymb,graphicx,subcaption,wrapfig,algorithm,algpseudocode}
\usepackage[margin=2cm]{geometry}

\begin{document}
\title{\bf \Large Group Testing for COVID-19: How to Stop Worrying and Test More}
\author{Lakshmi Narasimhan Theagarajan\\
Indian Institute of Technology Palakkad}
\date{}
\maketitle

\linespread{1.2}

\begin{abstract}
The corona virus disease 2019 (COVID-19) caused by the novel corona virus has an exponential rate of infection.
COVID-19 is particularly notorious as the onset of symptoms in infected patients are usually delayed and there exists a large number of asymptomatic carriers.
In order to prevent overwhelming of medical facilities and large fatality rate, early stage testing and diagnosis are key requirements. 
In this article, we discuss the methodologies from the group testing literature and its relevance to COVID-19 diagnosis.
Specifically, we investigate the efficiency of group testing using polymerase chain reaction (PCR) for COVID-19.
Group testing is a method in which multiple samples are pooled together in groups and fewer tests are performed on these groups to discern all the infected samples.
We study the effect of dilution due to pooling in group testing and show that group tests can perform well even in the presence of dilution effects.
We present multiple group testing algorithms that could reduce the number of tests performed for COVID-19 diagnosis.
We analyze the efficiency of these tests and provide insights on their practical relevance.
With the use of algorithms described here, test plans can be developed that can enable testing centers to increase the number of diagnosis performed without increasing the number of PCR tests.
The codes for generating test plans are available online at \cite{website}.  
\end{abstract}

\section{Introduction}
The novel corona virus has caused a pandemic in early 2020. 
The reproduction rate of the corona virus is estimated to be between 1.4 and 3.9 \cite{riou2020pattern}.
From an infected patient, COVID spreads to 1.4 to 3.9 others on an average. 
As the virus spreads through respiratory droplets, it spreads at an exponential rate, especially, in densely populated locations. 
The infected patients become carriers at an early stage even before the onset of symptoms \cite{bar2020sars}.
Thus, it becomes imperative to test a large number of people and identify early stage infections to contain the spread of COVID-19.
The PCR based testing is considered to be one of the best techniques for early stage diagnosis 
\cite{sheridan2020coronavirus}.
Group testing is well known methodology in the combinatorics and compressed sensing literature \cite{gilbert2008group}. 
In group testing(GT), multiple samples are pooled together to form groups (fewer in number than the total number of samples).
Tests are performed on these groups and from the outcome of these tests, the infected  samples are inferred.
Group testing relies on the fact that not all of the tested samples may be infected.
Thus, {\em by employing group testing for COVID-19 diagnosis, the number of samples tested can be increased, the tests performed can be made economical and the speed of detection can be improved}.
This can lead to efficient containment of the spread of COVID-19.

In this article, we investigate group testing for COVID diagnosis using PCRs from a practical view.
We analytically study the effect of dilution, caused by pooling in group tests, on the accuracy of the tests.
We present multiple group testing algorithms with which a test plan for diagnosis of a pool of samples can be generated.
We analyze and present the scenarios in which the presented algorithms are appropriate.
We show through simulations that group testing can provide considerable gains in the number of tests performed.
Finally, we provide some guidelines for performing COVID diagnosis in practice.
A practitioner can directly generate the test plans using the MATLAB codes that are available online at \cite{website}.

\noindent Certain key findings and discussions of this article can be listed as follows:
\begin{itemize}
\item When the infection rate is less than 33\%, group test can help to reduce the number of tests significantly. Some examples on the gains obtained through group testing (with a sensitivity of more than 99\%) are given below.
\begin{center}
\begin{tabular}{ |c|| c| c| c| c| c| c|}
\hline
Number of samples to test & 16 & 16 & 16 & 32 & 32 & 32\\\hline
Infection rate & 5\% & 10\% & 20\%& 5\% & 10\% & 20\% \\\hline
Number of tests required by group testing & 6 & 8 & 12 & 11 & 16 & 26\\\hline 
\end{tabular}
\end{center}

\item Replication or repeating a test is required to improve the accuracy of any qRT-PCR test outcome. 
We find that replicating the test twice could be sufficient to get high accuracy.
Further, group testing could provide inherent replication and reduce the number of replicates required.
For details refer to Sections \ref{sec31} and \ref{sec53}.

\item We also analyzed the effect of dilution caused by dividing a swab sample's content to smaller portions and mixing multiple samples together. 
This helped us to determine an upper limit on the number of samples that can be pooled together.
It was found that up to 57 samples can be pooled together without reducing the test accuracy.
For details refer to Section \ref{sec52}.

\item It was found that different group testing strategies are optimal under different conditions. 
The diagnostician should adapt the test plan depending on the infection rate of a given cluster or local community. 
Test plans that are optimal for a given infection rate are discussed in Section \ref{sec6}. 
\end{itemize}
Further guidelines to follow while testing for COVID-19 through group testing are discussed in Section \ref{sec7}.

{\em Notations}: $\lfloor.\rfloor$ denotes the floor operation, i.e., largest integer lesser than or equal to the given number. 
$\lceil.\rceil$ denotes the ceil operation, i.e., smallest integer greater than or equal to the given number. 
$\lceil.\rfloor$ denotes rounding off operation.
$|\mathcal{A}|$ denotes the cardinality of the set $\mathcal{A}$, i.e., the total number of elements in the set $\mathcal{A}$.

\section{Real-Time Reverse Transcription-PCR}
\begin{figure}[h]
\centering   
\includegraphics[width=14cm]{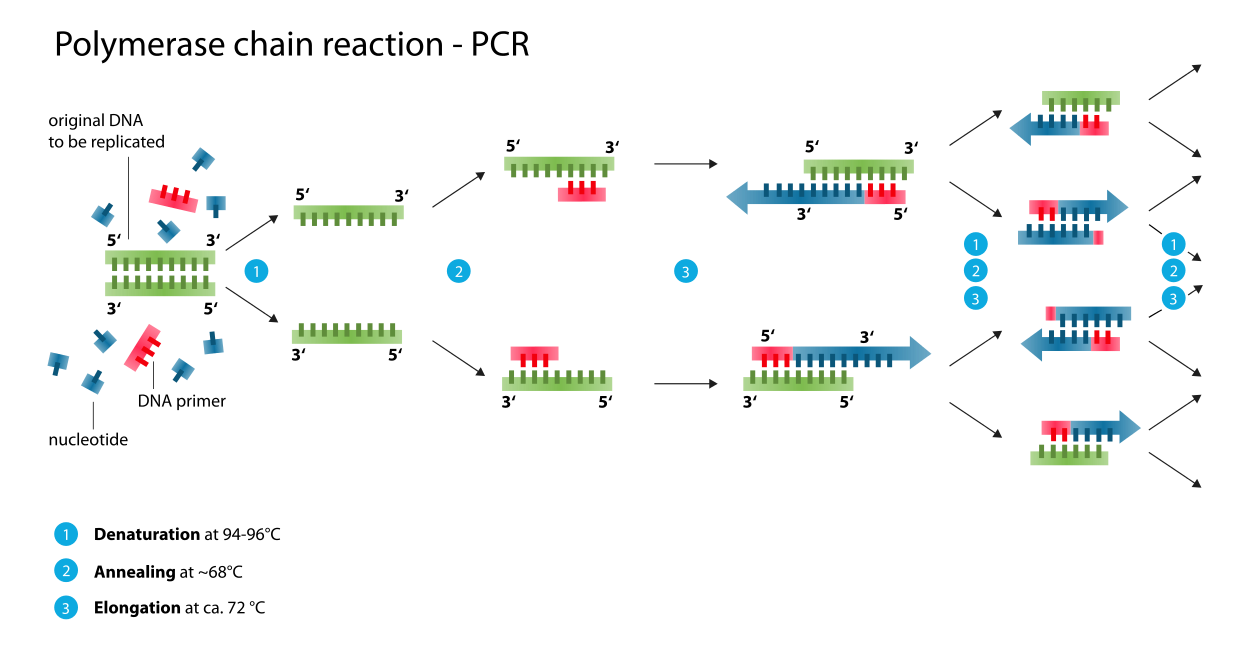}
\caption{\label{pcr}
Illustration of polymerase chain reaction (PCR). {\em Source: en.wikipedia.org/wiki/Polymerase\_chain\_reaction}
}
\end{figure}

The real time or quantitative reverse transcription polymerase chain reaction (qRT-PCR) is the testing method used for early detection of COVID.
In RT-PCR, the viral RNA molecules present in the test medium are first reverse transcribed into DNA, which is referred to as the complementary DNA (cDNA). 
The cDNA corresponding to the COVID genes are amplified using PCR, which is performed through thermal cycling. 
In the high temperature cycle, the target cDNA's double-helix strands are separated (referred to as the denaturation process).
In the low temperature cycle, the primers bind onto the cDNA strands (referred to as annealing) to initiate the polymerization through which the free nucleotides assemble on the DNA strands to form a new double-helix DNA (referred to as the elongation process).
Thus, in every cycle the number of DNA particles present are doubled; refer Fig. \ref{pcr}. 
After several cycles, the total number of target DNA present is increased to an exponentially large volume.
Finally, fluorescent dyes that fluoresce in the presence of the target DNA are used to visually identify the presence or absence of the virus.
The number of thermal cycles determine the final volume of target DNA present; hence, by increasing the thermal cycles (also referred to as amplification cycle), the sensitivity or the detection accuracy of qRT-PCR can be improved.
A detailed procedure and reagents used in qRT-PCR for detection of COVID can be found in \cite{corman2020diagnostic}.

\section{Goodness of Test}
\label{sec3}
Using qRT-PCR, it was found that 10 virus particles are sufficient to successfully detect the presence of the virus in a test sample with very high accuracy ($>$99.9\%) \cite{corman2020detection}. 
However, this accuracy or goodness of test may vary with the qRT-PCR kit used. 
We define the following metrics to compare and analyze the goodness of tests. 

Let $X$ be a binary random variable that denotes the presence ($X=1$) or absence ($X=0$) of the virus in a test sample.
Let $X_t$ be a binary random variable that denotes the outcome of a test.\\

\noindent{\bf Definition} - {\em False negative rate} : $\gamma = \Pr(X_t=0|X=1)$\\
The rate or probability of the outcome of a test being negative when the virus is actually present in the sample. 
Alternatively, the {\em sensitivity} is defined as the probability of the test outcome being positive when the virus is actually present in the sample; this is given by $1-\gamma = \Pr(X_t=1|X=1)$. 
It is desirable for the test to have a low value for $\gamma$.\\

\noindent{\bf Definition} - {\em False positive rate} : $\beta = \Pr(X_t=1|X=0)$\\
The rate or probability of the outcome of a test being positive when the virus is actually absent in the sample.
It is desirable for the test to have a low value for $\beta$.\\

\noindent{\bf Definition} - {\em Prior} : $\alpha = \Pr(X=1)$\\
The rate of occurrence of the virus in a given population.\\   

The prior can be approximately computed, using previous history, as the ratio of the number of positive cases to the total number of tests performed. 
It is important to minimize sample bias in this heuristic computation.
According to \cite{roser2020coronavirus}, the prior (at the time of writing this article) for United States of America is about 0.1973 and for India is about 0.0379.
Among the countries where more than $10^5$ tests were performed, the lowest prior is for UAE (about 0.005) and the highest is for Spain (about 0.4423).
In real life, we are more concerned about the false negative rate (as opposed to false positive rate). 
It is very important to have the least $\gamma$ to minimize the spread of COVID and fatality caused by it.
Further, the value of these rates depend on the viral load or the number of viral RNA copies present in the test (denoted by $l$).
We denote the false negative and positive rates as a function of $l$ using $\gamma(l)$ and $\beta(l)$, respectively.
It was found in \cite{corman2020detection} that $\gamma(5)\approx 0.07$. 

Therefore, the goodness or efficiency of a testing technique is given by $\gamma(l)$ and $\beta(l)$.
It is necessary to know these parameters for a given testing technique before we can analyze the efficiency of group testing using the same.     

\begin{figure}[b]
%\centering
\begin{subfigure}[t]{0.5\textwidth}
\includegraphics[width=7.5cm]{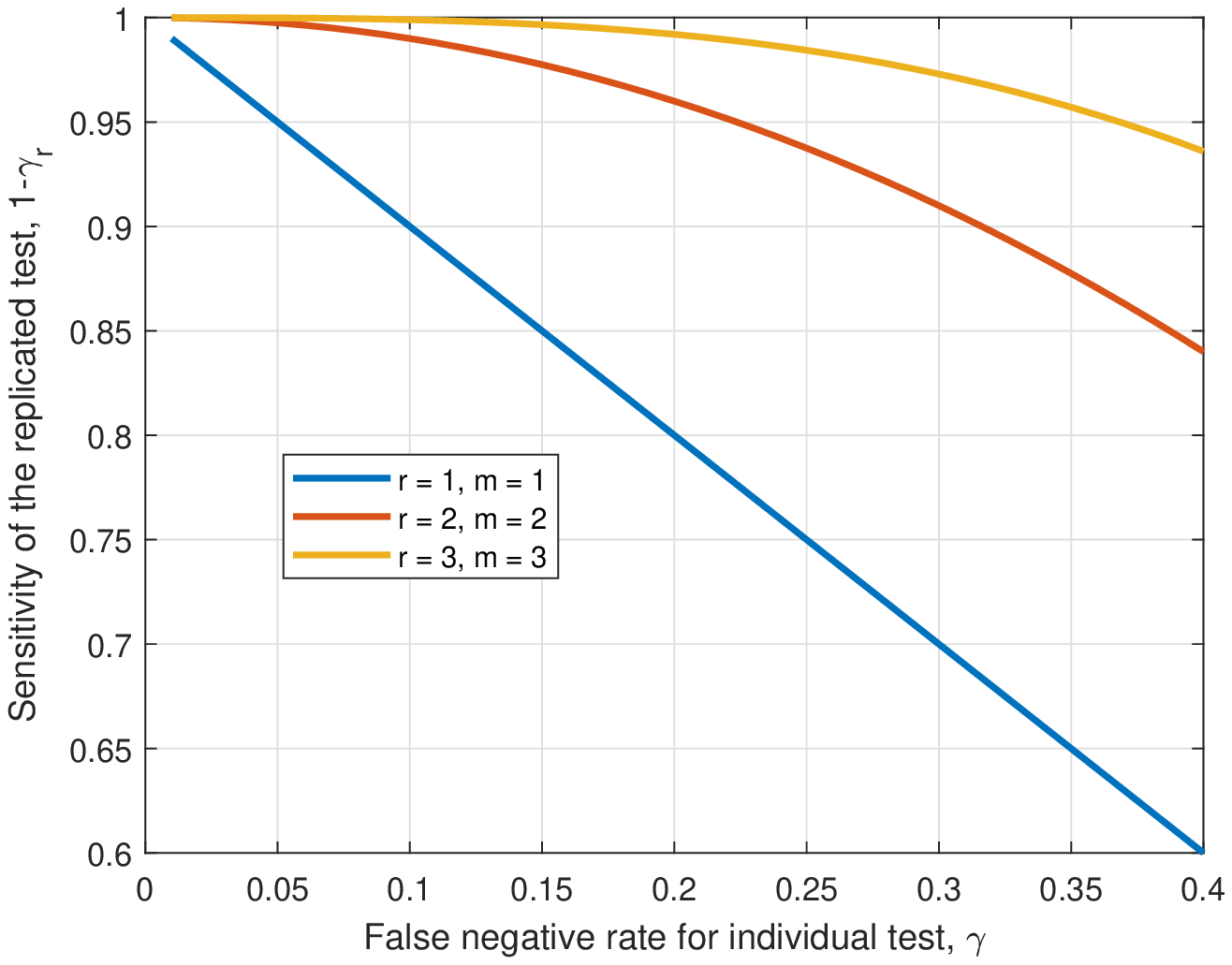}
\caption{\label{rep1} Variation of sensitivity for different replicates.}
\end{subfigure}
\begin{subfigure}[t]{0.5\textwidth}
\includegraphics[width=7.5cm]{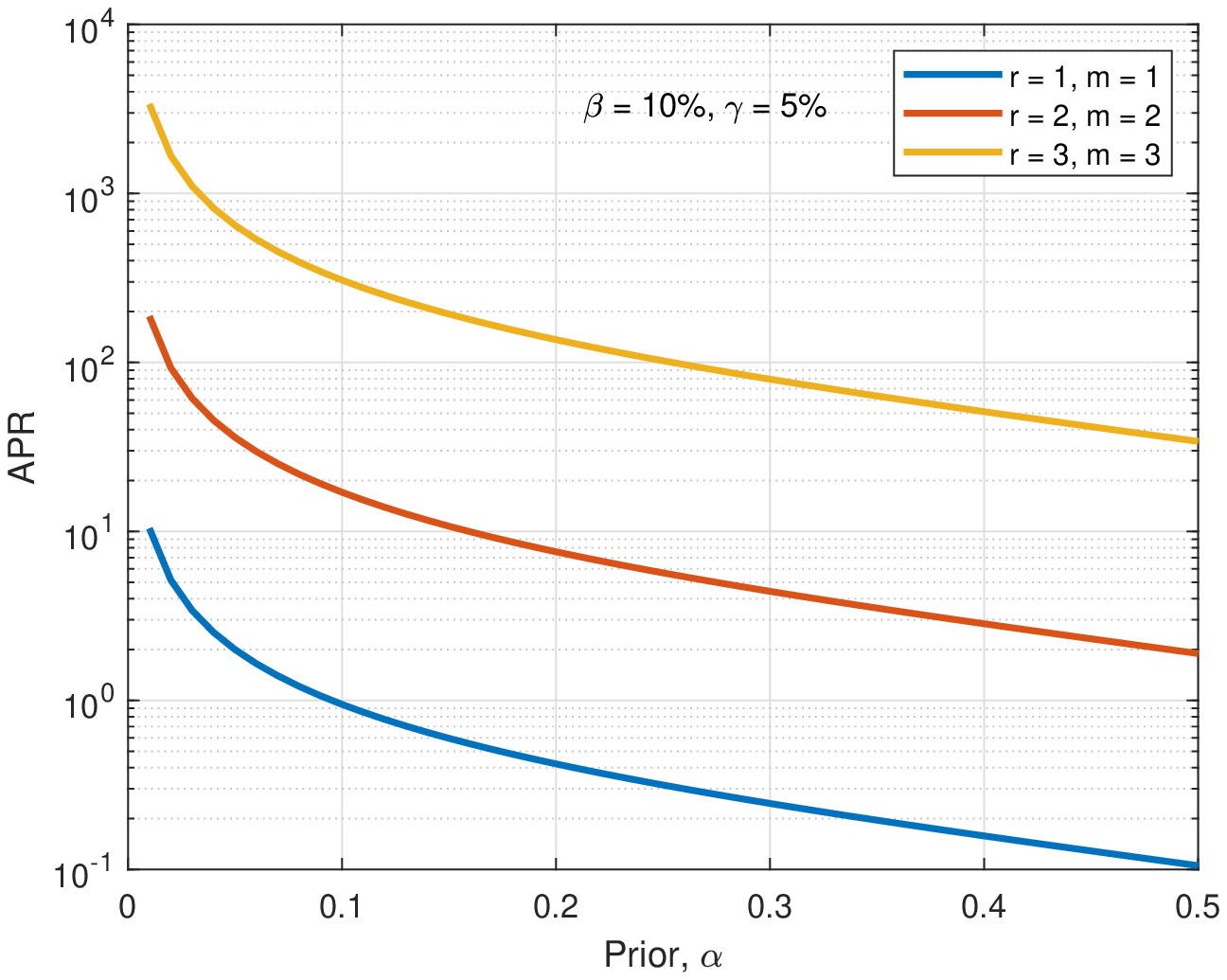}
\caption{\label{rep2} Variation of APR for different replicates.}
\end{subfigure}

\caption{\label{rep}
Effect of replication on the accuracy of the tests for $r$ replicates and $m$ number of negative outcomes.  
}
\end{figure}

\subsection{Replication}\label{sec31}
Often tests are performed multiple times on a single sample to confirm the outcome and account for any variability in the testing procedure, thereby improving the accuracy of the test.
Here, we shall analyze the improvement in accuracy when tests are replicated for a given sample. 

When the test is replicated $r$ times, the outcome is declared based on the majority rule.
To declare the final decision as negative, the APR should have a value greater than 1.  
The expressions to compute the sensitivity and APR due to replication are derived in Appendix \ref{ap1}.
Figure \ref{rep} illustrates the improvement in accuracy for double and triple replication. 
From Fig. \ref{rep1}, one can choose the value of $r$, i.e., how many tests to perform depending on the $\gamma$ of the testing technique and the $\gamma_r$ that is desired. 
For example, when $\gamma$ is small $(<0.05)$, double and triple replicates give very similar performance; hence, double replication will suffice.
Figure \ref{rep2} shows us that replication is necessary when the prior is more than 8\% at a false negative and positive rates of 5\% and 10\%, respectively, for the testing technique.

\section{Group Testing}\label{sec4}
Group testing(GT) refers to the idea of pooling multiple samples together and performing tests on certain subsets of these samples to discern the infected samples.

\begin{wrapfigure}{r}{0.625\linewidth}
\centering   
\includegraphics[width=11cm]{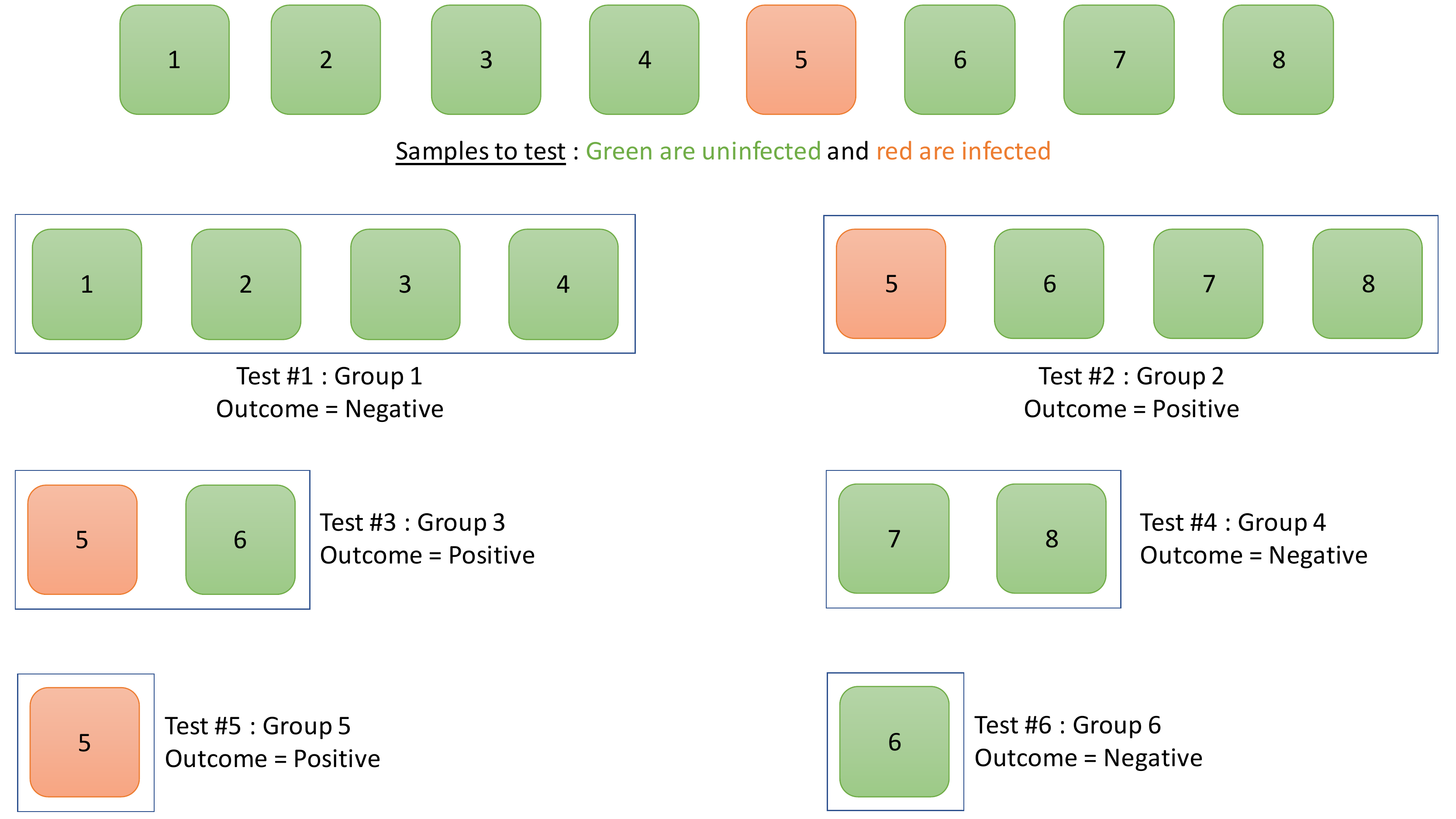}
\caption{\label{toy1}
Illustration of group testing with a simple example.  
}
\end{wrapfigure}
First, we illustrate GT through a simple example given in Fig. \ref{toy1}.
Let there be 8 samples to test and only one of it be infected. 
Pool samples \{1,2,3,4\} and \{5,6,7,8\} in two groups and test them. 
Discard group-1 that tests negative and carry forward group-2 that tests positive.
Form groups 3 and 4 with samples \{5,6\} and \{7,8\}, respectively; repeat as before. 
Here, we have utilized only 6 tests to diagnose 8 samples - a 25\% reduction in the number of tests.
Note that, if we had known that there was exactly one infected sample, then only 4 tests would be required (tests 4 and 6 would not be required).
This GT algorithm is referred to as the {\em binary search} and is the optimal GT when there is exactly one infected sample in a given pool.
In general, to test a pool of $N$ samples with up to one infected sample, we require at most $2\log_2N$ tests and at least $\log_2N$ tests. 
This gain increases exponentially with $N$; for example, $N=64$ requires at most 14 tests with GT as opposed to 64 tests without GT.  

The theory of group testing deals with the design and analysis of algorithms that tell us how to choose subsets (groups) of samples to pool together and test, and identify the infected samples.
In other words, GT aims to minimize the number of tests ($T$) required to identify at most $D$ number of infected samples among $N$ number of given samples (pool size); $0\leq D<N$ and $T\leq N$. 
In practice, the value of $D$ is not known. 
Nevertheless, depending on the value of the prior, an upper bound on the number of infected samples can be chosen\footnote{Note that it is important to randomize the samples when $D$ is chosen as a pessimistic estimate using the prior.} as $D$ for a given $N$.     

\subsection{Types of Group Tests}
There are two paradigms of GT, namely, 
\begin{itemize}
\item {\bf Combinatorial group testing} (CGT): 
The combinatorial algorithms require the exact number (or an upper bound) of the infected samples $D$. 
As long as the chosen value of $D$ is greater than or equal to the true number of infected samples, CGT algorithms always identifies the infected samples correctly.
% and the total number of tests performed in any scenario is bounded above by $N$.

\item {\bf Probabilistic group testing} (PGT): 
The probabilistic algorithms require an upper bound on $\alpha$ and identify all infected samples with certain probability $P_D$.
Usually, the detection probability $P_D$ is very close to 1; however, there still may be a non-zero probability of the infected cases going unidentified. 
For small values of $\alpha$, the average number of tests performed in PGT is less than $N$; however, there can exist cases for which the number of tests performed by PGT might be greater than $N$, albeit, the probability of occurrence of such cases will be relatively small.    
In PGT, there exists a trade-off between $P_D$ and the average value of $T$.
\end{itemize} 
From the above discussion, it can be seen that, due to their deterministic nature, CGT algorithms are preferred in practice to test for diseases.
When an upper bound on $D$ cannot be reliably established, PGT algorithms may prove to be more efficient as opposed to CGT.
There exists PGT algorithms in which $P_D=1$. 
In the worst case (i.e., when $D=N-1$), the value of $T$ can be greater than $N$, but the average value of $T$ can still be much less than $N$.
This is further discussed in Section \ref{sec6}. 
CGT and PGT algorithms are further classified into  
\begin{itemize}
\item {\bf Adaptive tests}: 
Here, the tests are performed sequentially. 
First a group is chosen randomly based on $\frac{D}{N}$ or $\alpha$ and tested, the outcome of this test determines the next group to test and so on.
Thus, the size and samples of a group are chosen adaptively based on previous group and its test outcome.
The GT described in Figure \ref{toy1} is an example of adaptive CGT.

\item {\bf Non-adaptive tests}:
When the test plan is fixed for a given $D$ and $N$, then it is known as the non-adaptive GT.
Here, a fixed number of tests are always performed irrespective of the number of infected samples present in the pool. 
An advantage of non-adaptive GT is that, if $T_N$ is the number of groups to be tested, then all the $T_N$ tests can be simultaneously run in parallel.
\end{itemize}
The non-adaptive GT can be represented by a matrix of dimension $T_N\times N$ (referred to as the measurement matrix), and the samples can be represented by a $N\times 1$ vector with 0's for uninfected samples and 1's for infected samples. 
The measurement matrix consists of 0's and 1's, the 1's in a row correspond to the samples included in a test and the 1's in a column correspond to the number of times a particular sample is tested.
The Boolean OR operation between the measurement matrix and the samples vector gives the GT outcomes.
Now, designing a non-adaptive GT algorithm becomes a problem of designing efficient measurement matrices\footnote{Constraints such as fixed row and column weights can be imposed in the design of measurement matrices.} and decoding algorithms to discern the positions of 1's in the sparse samples vector with the given $T_N$ test measurements.
The non-adaptive test algorithms can be studied with the aid of compressed sensing literature \cite{atia2012boolean}, \cite{gilbert2008group}. 
%For further information, refer \cite{gilbert2008group}. 

A key drawback of the non-adaptive GT algorithms is that it requires a large value of $N$ (in the order of $10^3-10^6$) to be efficient and provide reliable performance. 
Also, when the number of testing kits are limited and the prior is small, sequential tests are more efficient and quicker in identifying and discarding uninfected samples than parallel tests. 
Therefore, in this article, we shall limit our study only to adaptive tests.

%In the next section, we describe few pertinent algorithms that can be employed when $D$ is known and in the case where it is unknown.  

\section{Practical Considerations in Performing Group Tests for COVID}\label{sec5}
In this section, we shall discuss some limitations and surprising advantages of performing group tests for COVID diagnosis in practice using qRT-PCR.

\subsection{When is GT Efficient?}
A key question in the application of GT for COVID testing is how small should $D$ be, relative to $N$? 
That is, what is the range of values of $\frac{D}{N}$ for which GT reduces the total number of tests required?
This was first answered in \cite{fischer1999cut}; GT is said to reduce the number of tests required when $\frac{D}{N}< \frac{3-\sqrt{5}}{2}$.
As a general rule of thumb, \textmd{\em considerable reduction in the total number of tests can be achieved when $\frac{D}{N}$ or the prior $\alpha$ is less than 33\%}; in all other cases, it is best to perform individual tests.
As discussed in Section \ref{sec3}, typical value of $\alpha$ for COVID is less than 0.33.
Thus, GT can be utilized to efficiently increase the number of samples tested for COVID with minimal  number of tests performed.
 
\subsection{Effect of Dilution}\label{sec52}
In group testing, dilution occurs in two stages, namely, preparation and pooling. 
We shall discuss the effects of these dilutions on the allowable sample size and goodness of test.
The following discussion will also help us to determine a practically appropriate value of the pool size $N$ that provides reliable test outcomes.

\subsubsection{Sample Preparation}
When preparing the samples for GT, each sample needs to be divided into multiple portions for usage in multiple groups.   
In the example in Fig. \ref{toy1}, each sample is required to be divided into 3 portions since any sample could be involved in a maximum of 3 groups. 
In general, a sample may be involved in at most $\log_2N$ groups\footnote{The maximum number of groups in which a sample will be involved may vary depending on the GT algorithms. 
It can be proved that $\log_2N$ are sufficient to identify an infected sample among an arbitrary number of infected samples \cite{du2000combinatorial}.  
Further, once identified, as infected or uninfected, that sample is not used in any subsequent groups for testing.} while testing $N$ samples \cite{du2000combinatorial}.
If each test is replicated $r$ times, then a sample may have to be divided into $r\log_2N$ portions.
However, we shall see in the next subsection that each test need not be replicated in GT as it could provide inherent replication.

\begin{wrapfigure}{r}{0.5\linewidth}
\centering   
\includegraphics[width=8cm]{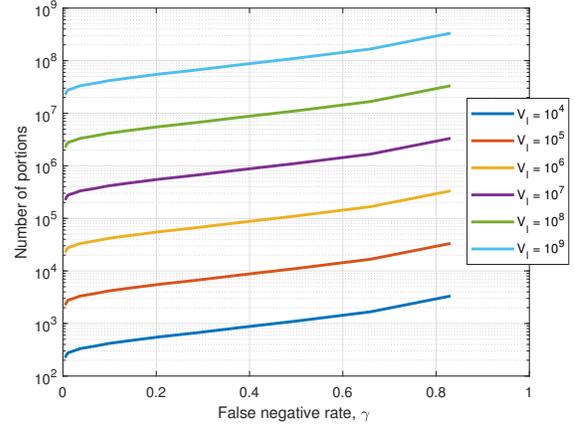}
\caption{\label{portions}
False negative rates achieved by different portion size at different sample viral loads ($V_l$).}
\end{wrapfigure}
In practice, to test for COVID, swabs from nasopharynx or throat \cite{bar2020sars} are taken.
The swab samples can be dissolved in liquid buffer media \cite{johnson1990transport} and this can be divided into multiple portions.  
When the samples are divided into multiple portions, the viral load gets distributed across the divided portions. 
Therefore, each sample can be divided only into certain number of portions such that the viral load in each portion can still be detected reliably by qRT-PCR.
The swab from nasopharynx contains $10^6 - 10^9$ corona viral particles \cite{bar2020sars}, if $L$ is the amount of viral load required for reliable detection, then each sample can be divided into at most $\frac{V_l}{L}$ portions, where $V_l$ is the amount of viral particles in the swab.
As discussed in Section \ref{sec3}, the false negative rate $\gamma(l)$ depends on the viral load in the test sample. 
If $\gamma^*$ is the required false negative rate for each test, then the corresponding viral load is given by the inverse function $\gamma^{-1}(\gamma*)$, and each sample can be divided into at most $\frac{V_l}{\gamma^{-1}(\gamma^*)}$.    
In Figure \ref{portions}, using the $\gamma(l)$ from \cite{corman2020detection} for qRT-PCR, we plot the number of portions into which a swab containing $V_l$ viral load can be divided to achieve a given false negative rate with three replicates.
It can be seen that, when the swab has a viral load of $10^4$, we can still obtain about 220 portions for $\gamma=0.02$. %a very low false negative rate. 

%Example: For a testing technique with $\gamma^{-1}(0.98)=10^4$, triple replication and a swab of viral load $10^6$, the pool size\footnote{The worst case number of tests in which a sample can be involved is the pool size $N$ itself.} can be a maximum of $\frac{V_l}{r\gamma^{-1}(0.98)}\approx 33$.
  
\begin{wrapfigure}{r}{0.5\linewidth}
\centering   
\vspace{-8mm}
\includegraphics[width=8cm]{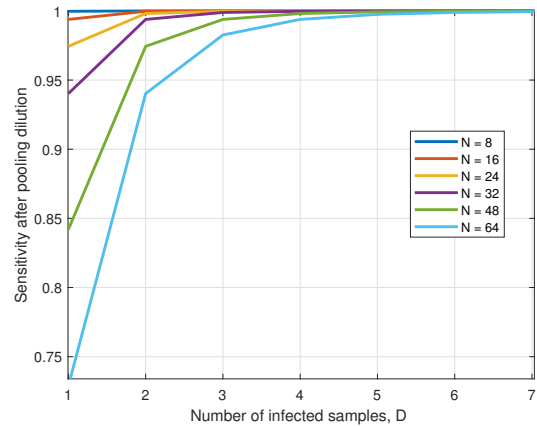}
\caption{\label{pd}
Effect of pooling dilution on sensitivity.}
\end{wrapfigure}
\subsubsection{Sample Pooling}
When portions of $N$ swabs, of which $D$ are infected, are mixed or pooled together for a qRT-PCR test, the viral load in $D$ samples are diluted by the $N-D$ virus-free samples. 
This dilution is referred to as the pooling dilution. 
The effect of this dilution on the performance of virus detection has been studied in \cite{nguyen2019methodology}.
Based on the probit model described in \cite{weusten2011refinement}, the effect of pooling dilution  was derived in \cite{nguyen2019methodology}.
Though the model derived in \cite{nguyen2019methodology} is for testing HIV, the authors mention that the model is applicable to other viral tests. 
Simplifying the model in \cite{nguyen2019methodology} for COVID, the sensitivity after pooling dilution is given by
\begin{equation}
1-\gamma_N = \Phi\left(1.6449\frac{\log\left(\frac{\chi DV_p}{NV_{50}}\right)}{\log\left(\frac{V_{95}}{V_{50}}\right)}\right),
\label{pdeq}
\end{equation} 
\noindent where $\Phi(.)$ is the CDF of the normal distribution, $\chi$ is the number of RNA copies per viral particle ($\chi=1$ in the case of corona virus), $D$ is the number of infected samples, $V_p$ is the viral load in each infected sample, $V_{50}$ and $V_{95}$ are the viral loads corresponding to a sensitivity of 0.5 and 0.95, respectively, for the considered testing technique, i.e., $V_x=\gamma^{-1}(1-x)$. 
In Fig. \ref{pd}, we show the reduction in sensitivity of the pooled test when different number of samples ($N$) are pooled together with different number of infected samples ($D$).
It can be seen that the pool size of 32 still provides a very high sensitivity after pooling dilution even when a single sample is infected.  
A similar conclusion on the pool size was reported through experimentation in \cite{yelin2020evaluation}. 
In general, if $T$ is the number of tests for COVID that are to be performed on a swab with viral load $V_l$ with $r$ replicates to achieve a sensitivity of $1-\gamma^*$, then the maximum pool size can be derived as 
\begin{equation}
N=\frac{V_l}{rTV_{50}}10^{-\frac{\Phi^{-1}(1-\gamma^*)}{1.6449}\log\frac{V_{95}}{V_{50}}},
\end{equation}
To achieve a sensitivity of 95\%, the above equation can be simplified\footnote{The worst case dilution scenario is when $D=1$.} to $N=\frac{V_l}{rTV_{95}}$.
For $T=log_2N$, we get 
\begin{equation}
N=e^{W_0\left(\frac{V_l}{rV_{95}\log_2e}\right)},
\label{n1}
\end{equation}
where $W_0(.)$ is the Lambert W function. 
This formulation includes the effect of dilution in both the sample preparation and pooling stages.
%For $T=N$, i.e, the worst case GT scenario, we get 
%\begin{equation}
%N=\sqrt{\frac{V_l}{rV_{95}}}.
%\label{n2}
%\end{equation}

\noindent Example: For a swab containing a viral load of $10^6$ viral particles, and a testing technique that requires $10^3$ viral particles to achieve 95\% sensitivity, the maximum pool size for 95\% sensitivity of the group test should be 
%\begin{itemize}
%\item 
$N=57$.\\ 
%- according to \eqref{n1},
%\item $N=18$ - according to \eqref{n2}.
%\end{itemize}

As seen from the above example, an efficient group test which performs the least number of total tests ($T$) per sample can enable us to increase the pool size, which, in turn, increases the testing speed and cost.
Therefore, employing an efficient GT algorithm for COVID testing is the key component in pooling tests.
We shall discuss some practically relevant algorithms in Section \ref{sec6}.

\subsection{Inherent Replication in GT}\label{sec53}
In group testing, samples that belong to a group, which has a test outcome of positive, are often tested again in smaller groups. 
This ensures that multiple tests are performed for some samples. 
This can be observed in the example described in Fig. \ref{toy1}; sample 5 is involved in 3 tests that are positive.   
In GT, almost every sample that tests positive is replicated at least twice.
This inherent replication reduces drastically the false positive rates of GT without an explicit replication of the individual tests.
The exact amount of reduction of the false positive rate depends on the GT algorithm.
Note that the group tests whose outcome are negative may not be replicated in GT and may require explicit replication to reduce false negative rates.
Therefore, the false negative rates or the sensitivity would suffice to be an appropriate metric to evaluate group tests and one can focus on reducing $\gamma$ to improve GT.

\begin{wrapfigure}{r}{0.5\linewidth}
\centering   
\vspace{-1cm}
\includegraphics[width=8cm]{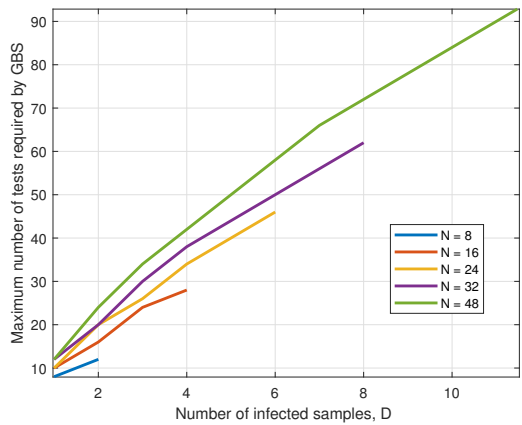}
\caption{\label{gbs}
 Worst case number of tests required by GBS with double replication for different values of $N$ and $D$.}
 \vspace{-1.5cm}
\end{wrapfigure}
\section{Algorithms}
\label{sec6}
In this section, we discuss two combinatorial group tests and a probabilistic group test which could be practically useful in testing for COVID-19.
The generalized binary splitting (GBS) and multi-stage testing (MST) are the CGT algorithms, and nested testing (NT) is the PGT algorithm that we describe in the following subsections.

\subsection{Generalized Binary Splitting}
The generalized binary splitting (GBS) is the most commonly used adaptive algorithm in the CGT literature \cite{du2000combinatorial}.
GBS is the generalization of the binary splitting procedure (BSP).
First, we describe BSP, and generalize it to obtain GBS.

The binary splitting procedure assumes that there exists at least one infected sample in a given pool and the goal of BSP is to identify exactly one infected sample in the least number of tests. 
That is, BSP works on a pool of size $N$ with $D\geq 1$ and identifies an infected sample in exactly $\log_2 N$ tests.

The steps in BSP are given below:
\begin{itemize}
\item {\bf Step 1} Split the $N$ samples into two halves, say groups $G_1$ and $G_2$. 
\item {\bf Step 2} Test $G_1$. 
\item {\bf Step 3} When the test is positive: (i) continue performing all future tests with the samples from only $G_1$, (ii) set $N=|G_1|$, and (iii) if the number of samples in $G_1$ is 1, then one infected sample has been identified and the algorithm is terminated.
\item {\bf Step 4} When the test is negative: (i) continue performing all future tests with the samples from only $G_2$, (ii) set $N=|G_2|$, and (iii) if the number of samples in $G_2$ is 1, then one infected sample has been identified and the algorithm is terminated.\\
Note that, since $D\geq1$, if $G_1$ tests negative, then $G_2$ must test positive.
\item {\bf Step 5} With the updated $N$ and samples pool, goto {\bf Step 1}.
\end{itemize} 

The GBS algorithm simply attempts to perform the BSP $D$ times to identify at most $D$ infected samples in a given pool of size $N$.
The pseudocode and the steps in GBS algorithm are given in Algorithm \ref{gbsalg}.\\
{\bf Analysis}: 
%The total number of tests performed ($T$) in GBS is at least $D\log_2N$. 
Since the outcome of a test performed at each step determines the next group for testing, the total number of tests varies for different inputs for a given $N$ and $D$.
The number of tests performed in the worst case, i.e., the maximum value that $T$ can attain in GBS is given by $T\leq \log_2{N \choose D}+D$. 
Figure \ref{gbs} shows the maximum number of tests required by GBS with each test replicated twice for different values of $N$ and $D$. 
It can be seen that GBS testing requires lesser number of tests, even in the worst case, than the conventional\footnote{Testing swabs one-by-one without any pooling or groups is referred to as the conventional testing.} testing, which is $2N$ . 
In GBS, each sample may be involved in up to $\log_2\frac{N}{D}+1$ tests; the sample preparation stage should provide at least $\log_2\frac{N}{D}+1$ portions for each sample.\\

\noindent{\bf Practical considerations}: When employing GBS in practice, the following points need to be kept in mind.
\begin{itemize}
\item The tests in GBS are sequential. 
That is, all the group tests have to be performed adaptively and in-order to obtain the final result. 
\item The gains of GBS are higher for large values of N. 
From simulations, we observed that GBS is suited for $\frac{D}{N}\leq 20\%$ and the average number of tests required can be reduced by up to 50\% of that of conventional tests. 
\item A key thing to note in GBS is that the algorithm can fail if the value of $D$ is underestimated. 
If the pool contains lesser number of infected samples than $D$, then GBS identifies all infected samples perfectly. 
\item {\em Inherent replication}: In GBS, when a group is tested positive, often it gets tested again in the subsequent groups, thereby providing inherent replication.
However, due to BSP, there exists scenarios where no samples are involved in multiple tests.
Therefore, in practice, one has to explicitly watch out for such samples and replicate the test, if required.   
\item At the end of GBS tests, the $N-D$ samples, that are marked as uninfected, can be pooled into one group and tested. This ensures that $D$ is not underestimated.
\end{itemize}
Some of the disadvantages of GBS are overcome by the multi-stage testing algorithm, which is described in the next subsection.

\subsection{Multi-stage Testing}
A disadvantage in GBS tests is that the number of group tests that a particular sample undergoes is difficult to compute. 
To overcome this issue, we can employ C. H. Li's multi-stage testing (MST) \cite{du2000combinatorial}.
In MST, each sample undergoes exactly $s$ number of tests.
The value of $s$ can be chosen as dictated by any practical restrictions; however, there is an optimal number of stages that would minimize the total number of tests performed, this is given by \cite{du2000combinatorial}
\begin{equation}
s = \underset{x\in\left\{\lfloor\ln\frac{N}{D}\rfloor,\lceil\ln\frac{N}{D}\rceil\right\}}{\arg\min}~~xD\left(\delta\right)^{\frac{1}{x}}, \qquad\qquad \text{ where } \quad \delta\triangleq\frac{N}{D}.
\end{equation}

The pseudocode of the MST algorithm is given in Algorithm \ref{mstalg}. 
It can be explained as follows. 
Maintain three bins: infected samples' bin (ISB), uninfected samples' bin (USB) and queued samples' bin (QSB).
Initially, all $N$ are in QSB. 
In the $i$th stage ($i=1,2,\cdots,s$), form $g_i$ groups with all samples in QSB and test them.
The samples that belong to groups which tested negative are moved to USB. 
If the group size is more than one, then the samples that belong to groups that tested positive are retained in QSB, else they are moved to ISB.
Proceed to $i+1$th stage and repeat the above steps till the group sizes become 1.  

\noindent{\bf Analysis}: The group size $g_i$ is computed as $g_i=\lceil N_{i-1}/k_i\rfloor$, where $k_i=\delta^{1-\frac{i}{s}}$ is the average number of samples in a group.
In practice, some groups of $g_i$ may contain $\lfloor k_i\rfloor$ samples and the rest may contain $\lfloor k_i\rfloor+1$ samples.
Note that, in the last stage, all groups contain exactly one sample, i.e., $k_s=1$.
The total number of tests performed in MST is $T=\sum_{i=1}^s g_i$.
The value of $T$ is not fixed for a given $N$ and $D$ as it depends on the group test outcomes.
However, we can determine the value of $T$ in the worst case, i.e., the maximum value that $T$ can take for  a given $N$ and $D$. 
This is given by $T\leq eD\ln\delta$ \cite{du2000combinatorial}.\\ 

\noindent{\bf Practical considerations}: When used in practice, MST has the following advantages.
\begin{itemize}
\item All groups in each stage can be tested in parallel. 
Thus, despite being an adaptive GT, testing can be sped up using MST.

\item When more than $D$ groups test positive at any stage, then it can be inferred that the estimate of $D$ is incorrect.
In this case, the remaining samples in QSB can be tested with an MST test plan for the updated value of $D$ and $N$ (the number of samples in QSB will be the new value of $N$).
This ensures that the tests performed in all previous stages are still useful even when the value of $D$ is incorrect.

\item When the true number of infected samples is less than or equal to $D$, MST identifies all infected samples.

\item Since the number of stages $s$ is known, the {\em number of times any sample will be tested is at most $s$}.
Thus, in the sample preparation, it is sufficient to divide each swab content to only $sr$ portions, where $r$ is the number of replicates. 

\item {\em Inherent replication}: Since all the samples in QSB at the end of stage-$i$ are tested again in stage-$i+1$, the tests for infected samples are automatically replicated.
Therefore, only those groups that test negative at each stage needs replication.
The groups that test positive are replicated inherently at least $s$ times, thereby increasing the test accuracy and reducing false positives.
This inherent replication further reduces the total number of tests required.
We derived that the total number of tests reduced due to inherent replication in MST is at least $k_1\frac{1-\frac{D}{N}}{1-\left(\frac{D}{N}\right)^{\frac{1}{s}}}$.

\item The maximum number of tests required by MST is always bounded above by $N$, i.e., $T\leq N$.
\end{itemize} 
\begin{figure}[h]
%\centering
\begin{subfigure}[t]{0.5\textwidth}
\includegraphics[width=9cm]{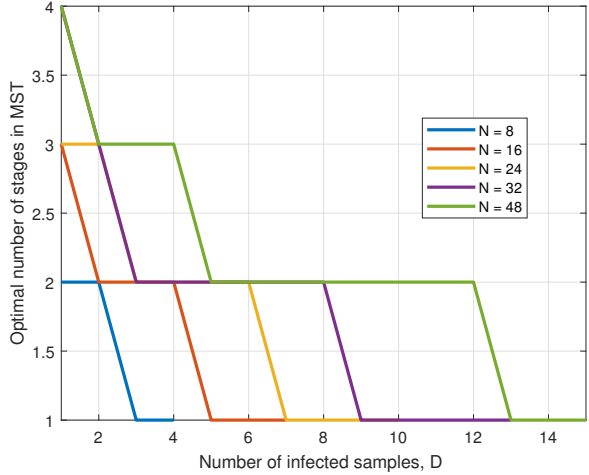}
\caption{\label{mstS} Optimal number of stages in MST.}
\end{subfigure}
\begin{subfigure}[t]{0.5\textwidth}
\includegraphics[width=9cm]{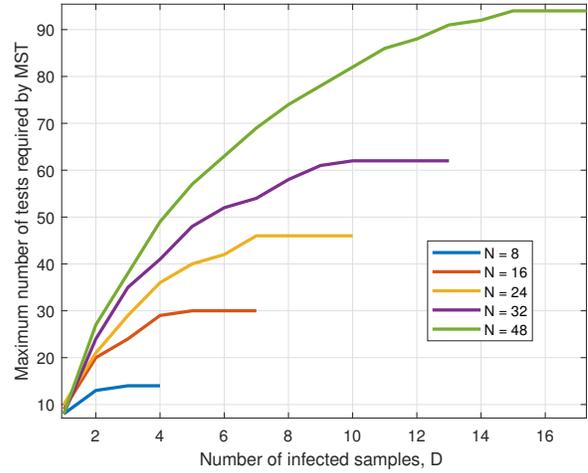}
\caption{\label{mstR} Worst case number of tests required by MST with double replication.}
\end{subfigure}
\caption{\label{mstf}
Optimal number of stages and the maximum value of $T$ (each test replicated twice) in MST for different values of $N$ and $D$.
}
\end{figure}
\begin{figure}
%\centering
\hspace{-1cm}\includegraphics[width=18cm]{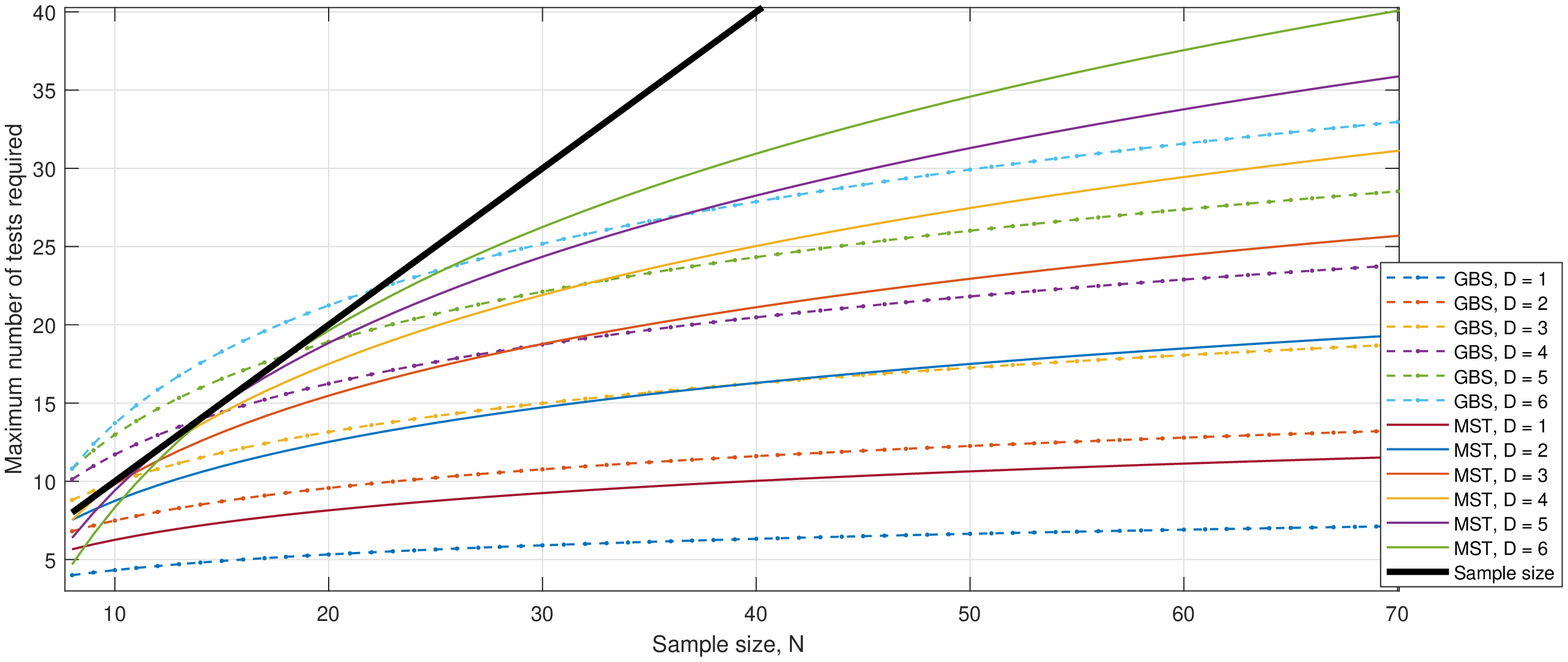}

\hspace{-1cm}\includegraphics[width=18cm]{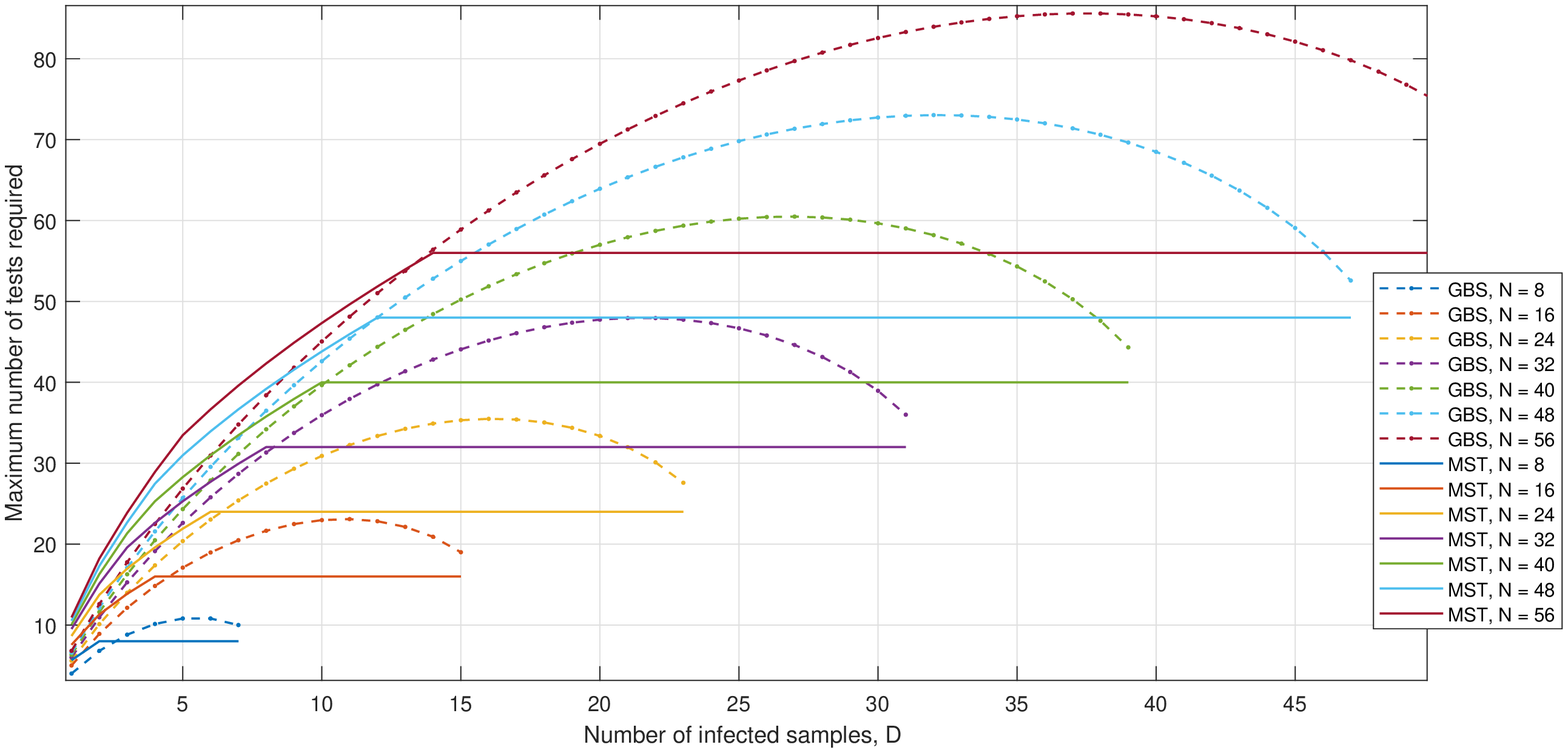}
\caption{\label{mstvsgbs} Comparison of the maximum number of tests required in group testing based on MST and GBS. The straight lines correspond to the number of tests required by conventional testing.}
\end{figure}

In Fig. \ref{mstS}, we plot the optimal number of stages required in MST for different values of $D$ and $N$.
Note that, when the number of stages increase, the inherent replication for infected samples increase and improves the accuracy without any additional tests for replication.
In Fig. \ref{mstR}, we plot the number of tests required by MST in the worst case scenario for different values of $D$ and $N$ with double replication.
For $N=16$, the maximum number of tests required for MST saturates at 30 as opposed to 32 in the conventional testing with double replicates.
It can be seen that MST requires uniformly lesser number of tests than the conventional testing.  
Though these numbers represent the worst case scenario, simulations show that the average number of tests required to detect infected samples could be 40\% less than the conventional testing.

\noindent{\bf GBS vs MST}: Figure \ref{mstvsgbs} shows the total number of tests required in the worst case for GBS and MST at different values of $N$ and $D=1,\cdots,6$. 
The number of tests required by conventional testing is also indicated for baseline comparison.
It can be seen that for small values of $\delta$, GBS outperforms MST.
As $D$ increases for a fixed $N$, GBS may sometimes require more number of tests than the conventional testing; however, MST always requires less than or equal number of tests as conventional testing.
In general, for large values of $N$ ($>$24), GBS may be appropriate, and MST may be appropriate in the other regime.
Figure \ref{mstvsgbs} can be used as a guideline to choose the best test for a given value of $N$ and $D$.

\subsection{Nested Testing}
The value of $D$ played an important role in the adaptive CGT algorithms discussed in the previous subsections.
When the value of $D$ is an underestimate, then the test plan provided by CGT may fail or may turn out to be suboptimal.
This shortcoming can be addressed by the usage of PGT algorithms.
Here, we shall describe a probabilistic group testing algorithm known as the nested testing (NT) \cite{wolf1985born}. 
The NT algorithm takes $N$ and $\alpha$ as the input and provides an adaptive test plan that minimizes the average number of total tests required to diagnose $N$ samples.
The actual number of infected samples in a pool or its estimate is not required for NT, only the prior probability of the presence of an infected sample is needed.
Nested testing is an optimal PGT which identifies all infected samples without failure \cite{wolf1985born}. 

\begin{wrapfigure}{r}{0.5\linewidth}
\centering   
\includegraphics[width=8cm]{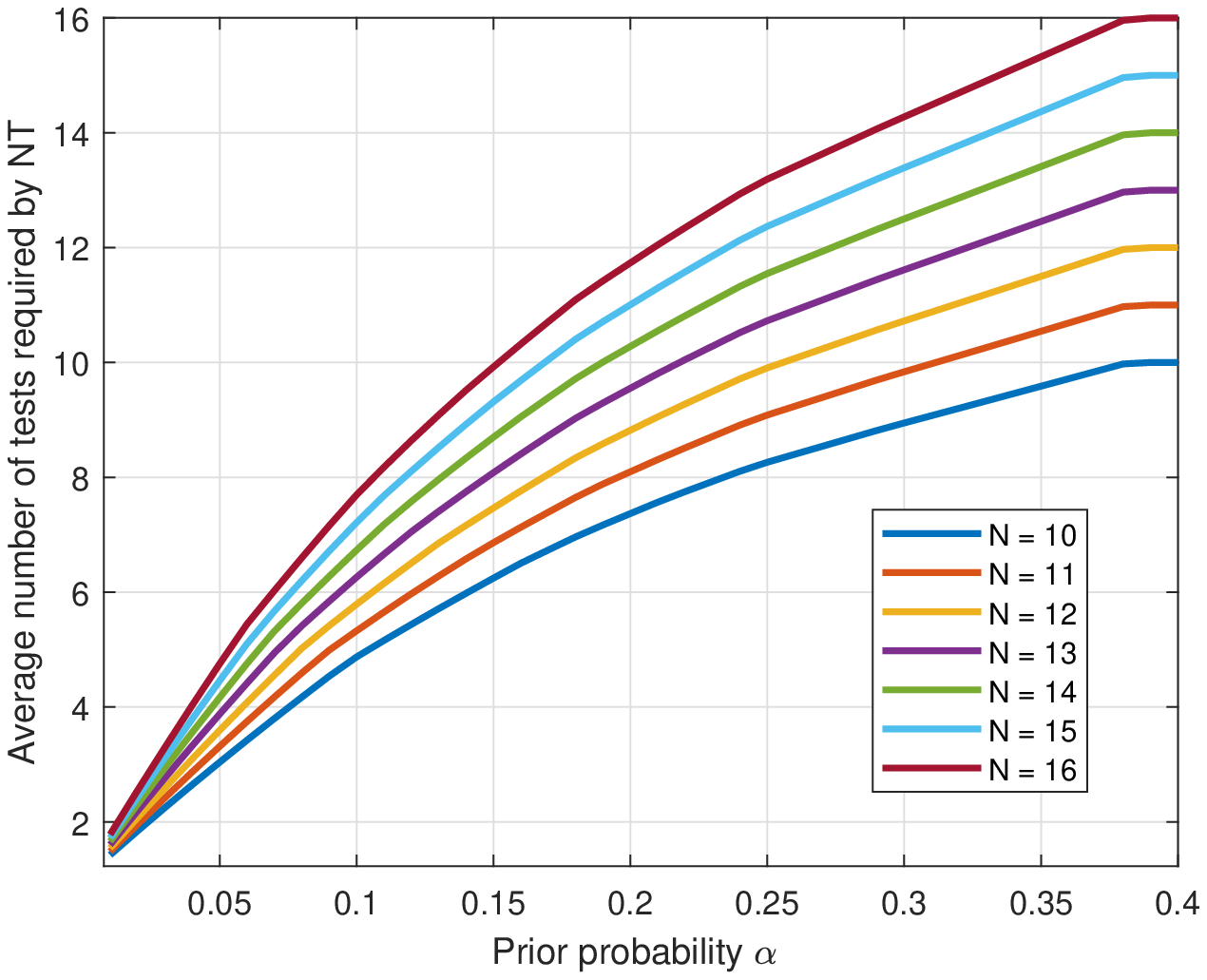}
\caption{\label{ntf}
Average number of tests required by nested testing for different prior probabilities.}
\end{wrapfigure}
The NT algorithm can be explained by considering three bins: 
\begin{enumerate}
\item Undiagnosed bin (UB) - this contains all the samples that are yet to be diagnosed, 
\item Potentially infected bin (PIB) - this contains a group of samples with at least one of them being an infected sample, and 
\item Diagnosed bin (DB) - this contains all the samples that are diagnosed as either positive or negative.
\end{enumerate}
The algorithm proceeds by testing a group of samples from UB and moving them to PIB, if they test positive, then transferring the diagnosed samples to DB and returning the rest to UB, and repeating the whole process till all samples are diagnosed.\\  
\noindent{\bf Analysis}: The NT algorithm models the presence of the virus in each sample as an independent Bernoulli random variable with probability $\alpha$.
Thus, NT can identify any number of infected samples in a pool, i.e., $0\leq D\leq N$.
The probability $\Pr(D=i), i=0,\cdots,N$, is given by the Binomial distribution; nested testing exploits this fact to identify the group that has the highest probability to contain infected samples and tests it.
The NT algorithm gives the least number of tests for which $\Pr(D=i)$ is the maximum. 
When the number of infected samples is $i$, such that $\Pr(D=i)$ is the least, the total number of tests required by NT may be more than $N$.
However, when the NT test plan is applied to multiple pools of size $N$, the average number of tests required per pool will be lesser than $N$.
The average number of tests required by NT is given by $G(0,N)$, where $G()$ is as given in \eqref{Geq}\footnote{The equations in \eqref{Geq} correspond to choosing the group size that maximize the probability of finding an infected sample that is binomially distributed in a group \cite{sobel1959group}.}.
Figure \ref{ntf} plots the average number of tests required by NT for different values of $N$ and $\alpha$.
It is clear that, for smaller values of the prior probability, NT can provide large gains in terms of the average number of tests performed without any knowledge of $D$. \\

\noindent{\bf Practical considerations}: When used in practice, NT has the following advantages.
\begin{itemize}
\item Nested testing can identify all infected samples irrespective of the actual value of $D$. 
The probability of detection of infected samples by NT is 1.

\item A key condition for NT to perform well in practice is the availability of independent samples.
That is, the probability of each sample being infected should be independent of the other samples in the pool and equal $\alpha$.
This can be achieved in practice through {\em randomization of samples}. That is, constituting a pool with samples from reasonably separated geographical locations. 

\item The value of the prior $\alpha$ can be estimated based on the past history as discussed in Section \ref{sec3}.
Further, this value should be continuously updated based on the outcomes of the tests performed each time.
 
\item {\em Inherent replication}: In NT, an infected sample has very high probability of getting tested in at least two groups.
Hence, NT can ensure at least double replication for samples that test negative.   

\item To create an NT test plan, the algorithm needs to evaluate the expression in \eqref{Geq}, this can be computationally intensive due to the recursive nature of the expressions.
Thankfully, the expressions can be computed offline for different values of $N$ and $\alpha$, and stored as a look-up table.
This look-up table can be used in the algorithm to speed up the testing in real-time.  
% table in sobel
\end{itemize}

\section{Guidelines to Employ Group Testing for COVID-19 diagnosis}\label{sec7}
Group Testing is a promising method that can be effectively utilized to ramp up the number of tests performed in COVID-19 diagnosis.
GT can help us quickly identify several early stage infections.
By reducing the number of tests performed, GT can make the testing less expensive and economical.
The following are some guidelines that can be followed in employing GT for COVID-19 diagnosis.

\begin{itemize}
\item Before employing pooling, the sensitivity of the testing technique needs to be characterized. 
Some qRT-PCR kits are known to provide accurate results with smaller viral load, whereas some are known to require a large amount of viral load to detect the virus.
This sensitivity characteristics, i.e., $\gamma(l)$, needs to be understood before the formulation of GT test plans.
The knowledge of $\gamma(l)$ would help us to decide the optimal pool size and number of replications required.

\item When the testing speed is a primary factor, it may be better to employ MST as it enables to perform parallel group tests. 

\item In CGT algorithms such as GBS or MST, randomization is not required.
The statistical relation between the samples are irrelevant in CGT group tests. 
However, the estimate of the upper bound on the number of infected samples in a pool is an important  parameter that needs to be accurate.
The value of $D$ can be estimated by observing the history of a location or cluster.
When more samples from a location test positive, then the estimate of $D$ should be correspondingly increased and vice versa.

\item PGT algorithms such as NT has the advantage of not requiring the exact value or an upper bound on the number of infected samples in a pool.
PGT algorithms require the prior probability values ($\alpha$). 
Once again, this can be obtained from the history of the tests performed for a location or cluster.
The ratio of the number of samples tested positive to the total number of tests performed can give us this estimate (refer Section \ref{sec3}).
This estimate needs to be update continuously over time.

\item Randomization could significantly reduce the number of tests performed in PGT.
When samples from a small community or cluster are pooled together, most of them are likely to have a similar test outcome. 
However, PGT works best when the samples are independent.
Hence, the testing center should randomize by picking samples from different communities or clusters to form a pool.
This randomization can bring down the average fraction of samples that are positive in a pool and the dependence among the samples.

\item In practice, PGT tests are suited well for locations where the infection rate is higher and the CGT tests are suited well for locations where the infection rate is small.
This is because, PGT does not need the value of $D$ and an underestimated value could cause failure of CGT tests.

\item When choosing a GT algorithm to perform diagnosis, the practitioner can intelligently switch between GBS, MST and NT, depending on the estimate of $D$ and $\alpha$, and their accuracy. 
The performance plots provided in Section \ref{sec6} can be utilized in choosing the best test for a given scenario.

\item From the study so far, it is clear that considerable reduction in the total number of tests can be achieved when $\frac{D}{N}$ or the prior $\alpha$ is less than 33\%.

\item Note that when $D> \frac{N}{2}$ or $\alpha>\frac{1}{2}$, group testing can still be helpful. 
Under this condition, the goal of GT would be reversed, i.e., to identify the uninfected samples rather than the infected samples. 
All our discussions so far remain applicable with this switch.   

\item The MATLAB codes for simulating the results and generating group test plans are available online at \cite{website}.

\end{itemize}

\section*{Acknowledgments}
The author thanks Namrata M. Nilavar, Indian Institute of Science, Bangalore, for useful discussions.

\bibliographystyle{IEEEtran}
\bibliography{groupTestingCovid19}

\begin{thebibliography}{10}
\providecommand{\url}[1]{#1}
\csname url@samestyle\endcsname
\providecommand{\newblock}{\relax}
\providecommand{\bibinfo}[2]{#2}
\providecommand{\BIBentrySTDinterwordspacing}{\spaceskip=0pt\relax}
\providecommand{\BIBentryALTinterwordstretchfactor}{4}
\providecommand{\BIBentryALTinterwordspacing}{\spaceskip=\fontdimen2\font plus
\BIBentryALTinterwordstretchfactor\fontdimen3\font minus
  \fontdimen4\font\relax}
\providecommand{\BIBforeignlanguage}[2]{{%
\expandafter\ifx\csname l@#1\endcsname\relax
\typeout{** WARNING: IEEEtran.bst: No hyphenation pattern has been}%
\typeout{** loaded for the language `#1'. Using the pattern for}%
\typeout{** the default language instead.}%
\else
\language=\csname l@#1\endcsname
\fi
#2}}
\providecommand{\BIBdecl}{\relax}
\BIBdecl

\bibitem{website}
L.~N. Theagarajan, ``Available online: Matlab codes for test plan generation,''
  \emph{https://sites.google.com/view/lakshminarasimhan/research}.

\bibitem{riou2020pattern}
J.~Riou and C.~L. Althaus, ``Pattern of early human-to-human transmission of
  wuhan 2019 novel coronavirus (2019-ncov), december 2019 to january 2020,''
  \emph{Eurosurveillance}, vol.~25, no.~4, 2020.

\bibitem{bar2020sars}
Y.~M. Bar-On, A.~Flamholz, R.~Phillips, and R.~Milo, ``Sars-cov-2 (covid-19) by
  the numbers,'' \emph{eLife}, vol.~9, p. e57309, 2020.

\bibitem{sheridan2020coronavirus}
C.~Sheridan, ``Coronavirus and the race to distribute reliable diagnostics,''
  \emph{Nat Biotechnol}, 2020.

\bibitem{gilbert2008group}
A.~C. Gilbert, M.~A. Iwen, and M.~J. Strauss, ``Group testing and sparse signal
  recovery,'' in \emph{2008 42nd Asilomar Conference on Signals, Systems and
  Computers}.\hskip 1em plus 0.5em minus 0.4em\relax IEEE, 2008, pp.
  1059--1063.

\bibitem{corman2020diagnostic}
V.~Corman, T.~Bleicker, S.~Br{\"u}nink, C.~Drosten, and M.~Zambon, ``Diagnostic
  detection of 2019-ncov by real-time rt-pcr,'' \emph{World Health
  Organization, Jan}, vol.~17, 2020.

\bibitem{corman2020detection}
V.~M. Corman, O.~Landt, M.~Kaiser, R.~Molenkamp, A.~Meijer, D.~K. Chu,
  T.~Bleicker, S.~Br{\"u}nink, J.~Schneider, M.~L. Schmidt \emph{et~al.},
  ``Detection of 2019 novel coronavirus (2019-ncov) by real-time rt-pcr,''
  \emph{Eurosurveillance}, vol.~25, no.~3, 2020.

\bibitem{roser2020coronavirus}
M.~Roser, H.~Ritchie, and E.~Ortiz-Ospina, ``Coronavirus disease
  (covid-19)--statistics and research,'' \emph{Our World in Data}, 2020.

\bibitem{atia2012boolean}
G.~K. Atia and V.~Saligrama, ``Boolean compressed sensing and noisy group
  testing,'' \emph{IEEE Transactions on Information Theory}, vol.~58, no.~3,
  pp. 1880--1901, 2012.

\bibitem{fischer1999cut}
P.~Fischer, N.~Klasner, and I.~Wegenera, ``On the cut-off point for
  combinatorial group testing,'' \emph{Discrete Applied Mathematics}, vol.~91,
  no. 1-3, pp. 83--92, 1999.

\bibitem{du2000combinatorial}
D.~Du, F.~K. Hwang, and F.~Hwang, \emph{Combinatorial group testing and its
  applications}.\hskip 1em plus 0.5em minus 0.4em\relax World Scientific, 2000,
  vol.~12.

\bibitem{johnson1990transport}
F.~B. Johnson, ``Transport of viral specimens.'' \emph{Clinical microbiology
  reviews}, vol.~3, no.~2, pp. 120--131, 1990.

\bibitem{nguyen2019methodology}
N.~T. Nguyen, H.~Aprahamian, E.~K. Bish, and D.~R. Bish, ``A methodology for
  deriving the sensitivity of pooled testing, based on viral load progression
  and pooling dilution,'' \emph{Journal of translational medicine}, vol.~17,
  no.~1, p. 252, 2019.

\bibitem{weusten2011refinement}
J.~Weusten, M.~Vermeulen, H.~van Drimmelen, and N.~Lelie, ``Refinement of a
  viral transmission risk model for blood donations in seroconversion window
  phase screened by nucleic acid testing in different pool sizes and repeat
  test algorithms,'' \emph{Transfusion}, vol.~51, no.~1, pp. 203--215, 2011.

\bibitem{yelin2020evaluation}
I.~Yelin, N.~Aharony, E.~Shaer-Tamar, A.~Argoetti, E.~Messer, D.~Berenbaum,
  E.~Shafran, A.~Kuzli, N.~Gandali, T.~Hashimshony \emph{et~al.}, ``Evaluation
  of covid-19 rt-qpcr test in multi-sample pools,'' \emph{medRxiv}, 2020.

\bibitem{wolf1985born}
J.~Wolf, ``Born again group testing: Multiaccess communications,'' \emph{IEEE
  Transactions on Information Theory}, vol.~31, no.~2, pp. 185--191, 1985.

\bibitem{sobel1959group}
M.~Sobel and P.~A. Groll, ``Group testing to eliminate efficiently all
  defectives in a binomial sample,'' \emph{Bell System Technical Journal},
  vol.~38, no.~5, pp. 1179--1252, 1959.

\end{thebibliography}

\appendix
\section{Replication analysis}
\label{ap1}
Hence, the false negative and positive rates for $r$ tests are given by (each replicate is assumed to be independent of the other) 
\begin{eqnarray}
\gamma_r\triangleq\Pr\left(\sum_{t=1}^rX_t< \left\lceil \frac{r}{2}\right\rceil\Bigg|X=1\right) = \sum_{t=\left\lfloor \frac{r}{2}\right\rfloor+1}^r{r \choose t}\gamma^t(1-\gamma)^{r-t},\\
\beta_r\triangleq\Pr\left(\sum_{t=1}^rX_t\geq \left\lceil \frac{r}{2}\right\rceil\Bigg|X=0\right) = \sum_{t=\left\lceil \frac{r}{2}\right\rceil}^r{r \choose t}\beta^t(1-\beta)^{r-t},
\end{eqnarray} 
where $X_t$ is the random variable denoting the outcome of the $t$th replicate. 
When the test is replicated twice, $\Pr(X_1=0, X_2=0|X=1) = \gamma^2$.
For triple replication, %$\Pr(\text{2 negatives + 1 positive}) = \gamma^2(1-\gamma)$ and 
$\Pr(X_1+X_2+X_3\leq 1|X=1) = 3\gamma^2(1-\gamma) + \gamma^3$ and $\Pr(X_1=0,X_2=0,X_3=0|X=1) = \gamma^3$.

For a given set of test replicate outcomes $X_1,\cdots,X_r$, the probability of the sample actually containing the virus is given by $\Pr(X=1|X_1,\cdots,X_r)$ - this is referred to as the {\em a posteriori} probability.
The a posteriori probability ratio (APR) $\frac{\Pr(X=0|X_1,\cdots,X_r)}{\Pr(X=1|X_1,\cdots,X_r)}$ is easier to compute than the individual a posteriori probabilities.   
The APR for a given set of outcomes is   
\begin{equation}
APR = \frac{\Pr(X=0|X_1,\cdots,X_r)}{\Pr(X=1|X_1,\cdots,X_r)}=\frac{\Pr(X_1,\cdots,X_r|X=0)}{\Pr(X_1,\cdots,X_r|X=1)}\frac{1-\alpha}{\alpha} = \frac{(1-\alpha)\beta^{r-m}(1-\beta)^{m}}{\alpha\gamma^{m}(1-\gamma)^{r-m}},
\end{equation}
where $m$ is the number of negative outcomes (i.e., $X_t=0$) in $r$ replicates.

\newpage
\section{Pseudocodes}

\begin{algorithm}
\caption{Multi-stage Testing}
\label{mstalg}
\begin{algorithmic}[1]
    \State {\bf Input}: $N, D, s$ and samples pool $\mathcal{P}_0=\{1,2,\cdots,N\}$.
	\State $N_0=N$
    \For {$i = 1 $ to $s$}
        \State $k_i=\delta^{1-\frac{i}{s}}$
		\State $g_i=\lceil N_{i-1}/k_i\rfloor$
		\State Divide $\mathcal{P}_{i-1}$ into $g_i$ disjoint groups - $\mathcal{G}_1,\mathcal{G}_2,\cdots,\mathcal{G}_{g_i}$ such that $\mathcal{G}_1\cup\mathcal{G}_2\cup\cdots\cup\mathcal{G}_{g_i}=\mathcal{P}_{i-1}$
		\State Test groups $\mathcal{G}_1,\mathcal{G}_2,\cdots,\mathcal{G}_{g_i}$
		\State Discard groups that tested negative
		\State $\mathcal{P}_i$ = \{samples from groups that tested positive\}, $\mathcal{P}_{i}\subset\mathcal{P}_{i-1}$ 	
		\State $N_i=|\mathcal{P}_i|$
	\EndFor
	\State Return $\mathcal{P}_{s}$ as the set of infected samples
\end{algorithmic}
\end{algorithm}

\begin{algorithm}
\caption{Generalized Binary Splitting Test}
\label{gbsalg}
\begin{algorithmic}[1]
    \State {\bf Input}: $N, D$ and samples pool $\mathcal{P}=\{1,2,\cdots,N\}$.
    \While {$N\geq2D-1$ and $D>0$}
        \State Choose a group $\mathcal{G}$ of size $2^{\lfloor\log_2\frac{N-D+1}{D}\rfloor}$
		\State Test group $\mathcal{G}\subseteq\mathcal{P}$
		\If {test outcome is positive}
		\State Identify an infected sample in $\mathcal{G}$ with BSP ({\scriptsize Since the group tested positive, it must contain at least one infected sample})
		\State Update $N=N-1-g$ ({\scriptsize where $g$ is the number of uninfected items diagnosed from BSP, remove these from the pool})
		\State Update $D=D-1$ ({\scriptsize remove the identified infected sample from the pool})
		\Else
		\State Update $N=N-|\mathcal{G}|$ ({\scriptsize $|\mathcal{G}|$ is the number of uninfected items in $\mathcal{G}$, remove these from the pool})
		\EndIf
	\EndWhile
	\If {$D>0$ and $N>0$}
	\State Test the $N$ samples individually
	\EndIf
\end{algorithmic}
\end{algorithm}

\begin{algorithm}
\caption{Nested Testing}
\label{ntalg}
\begin{algorithmic}[1]
    \State {\bf Input}: $N, \alpha$ and samples pool $\mathcal{U}=\{1,2,\cdots,N\}$.
	\State UB = $\mathcal{U}$; PIB = $\mathcal{P}$ and DB = $\mathcal{D}$ are empty
    \While { $\mathcal{U}$ is not empty}
		\State Compute  $h = \tilde{G}(0,|\mathcal{U}|)$ using \eqref{Gval}
		\State Test a group $\mathcal{G}\subseteq \mathcal{U}$ of size $h$
		\State Update $\mathcal{U}=\mathcal{U}-\mathcal{G}$
		\If {Test is negative }
			\State Update $\mathcal{D}=\mathcal{D}+\mathcal{G}$
		\Else
			\State Update $\mathcal{P}=\mathcal{G}$	
			\If {h == 1}
				\State Update $\mathcal{D}=\mathcal{D}+\mathcal{G}$ and make $\mathcal{P}$ empty
			\Else
			\While {$\mathcal{P}$ is not empty}
				\State Compute  $g = \tilde{G}(|\mathcal{P}|, |\mathcal{P}\cup\mathcal{U}|)$ using \eqref{Gval}
				\State Test a group $\mathcal{G}\subseteq \mathcal{P}$ of size $g$
				\If {Test is positive}
					\State Update $\mathcal{U}=\mathcal{U}+\mathcal{P}-\mathcal{G}$ and $\mathcal{P}=\mathcal{G}$
				\Else
					\State Update $\mathcal{D}=\mathcal{D}+\mathcal{G}$ and $\mathcal{P}=\mathcal{P}-\mathcal{G}$
				\EndIf
				\If {$|\mathcal{P}|$ == 1}
					\State Update $\mathcal{D}=\mathcal{D}+\mathcal{P}$ and make $\mathcal{P}$ empty
				\EndIf		
			\EndWhile
			\EndIf
		\EndIf
	\EndWhile
\end{algorithmic}
\begin{equation}
\tilde{G}(0,n) = \underset{x=1,\cdots,n}{\arg\min} \hat{G}(0,n,x) \qquad \text{ and }\qquad \tilde{G}(m,n) = \underset{x=1,\cdots,m-1}{\arg\min} \hat{G}(m,n,x),
\label{Gval}
\end{equation}
\begin{equation}
{G}(0,n) = \underset{x=1,\cdots,n}{\min} \hat{G}(0,n,x) \qquad \text{ and }\qquad {G}(m,n,x) = \underset{x=1,\cdots,m-1}{\min} \hat{G}(m,n,x),
\nonumber
\end{equation}
\begin{eqnarray}
\hat{G}(0,n,x) &=& 1 + (1-\alpha)^xG(0,n-x)+ (1-(1-\alpha)^x)G(x,n), \nonumber\\
\hat{G}(m,n,x) &=& 1 + \frac{(1-\alpha)^x-(1-\alpha)^m}{1-(1-\alpha)^m}G(m-x,n-x)+ \frac{(1-(1-\alpha)^x)}{1-(1-\alpha)^m}G(x,n), \label{Geq}\\
G(1,m) &=& G(0,m-1), \qquad G(0,1)=1, \qquad G(0,0)=0.\nonumber 
\end{eqnarray}
\end{algorithm}

\end{document}